\documentclass[groupedaddress,nofootinbib,prd,11pt]{revtex4}
\usepackage{graphicx}
\usepackage{cancel}
\usepackage{amssymb}
\usepackage{textcomp}
\usepackage{amsmath}
\usepackage{bm}
\usepackage{times}
\usepackage{color}

\begin{document}

\title{Behavior of perturbations on spherically symmetric backgrounds in multi-Galileon theory}
\author{Sebasti\'an Garc\'ia-S\'aenz}
\email{sgsaenz@phys.columbia.edu}
\affiliation{Department of Physics, Columbia University, New York, New York 10027, USA}

\begin{abstract}
We consider multi-Galileon theory, the most general Galilean invariant theory with $N$ scalar fields linearly coupled to the trace of the stress-energy tensor. We study the behavior of perturbations on a static spherically symmetric background with a massive point source, and show that, under the assumptions of stability and successful Vainshtein screening, solutions cannot be found that are free of both superluminal propagation and slowly moving, strongly coupled fluctuations. The latter imply that the theoretical and phenomenological issues related to a very low strong-interaction scale cannot be avoided in this model.
\end{abstract}
\maketitle
\section{Introduction}
Galileon theory was introduced as a local infrared modification of general relativity (GR) that could potentially explain the observed cosmic acceleration in a natural way, thanks to the existence of stable self-accelerating background solutions \cite{Nicolis:2008in}. Its origin can be traced back to the DGP model \cite{Dvali:2000hr}, a higher-dimensional brane-world model in which the graviton appearing in the 4-dimensional effective theory propagates additional degrees of freedom. It is possible to thoroughly study most of the interesting properties of this setup in the so-called decoupling limit \cite{Nicolis:2004qq}, in which the modifications to GR are encoded in a scalar field $\pi$ whose Lagrangian is invariant under the Galilean symmetry $\pi\to\pi+b_{\mu}x^{\mu}+c$. This motivated the study in Ref.\ \cite{Nicolis:2008in} of the most general theory exhibiting this Galilean invariance \cite{Nicolis:2010se} (with second-order equations of motion), providing a framework in which well-behaved self-accelerating solutions could be found \cite{Silva:2009km}.

The phenomenology of Galileon theories has proved to be extremely rich, and in the lapse of only a few years this class of theories has come to be recognized as an important alternative to other more usual modified gravity theories. Galilean invariance has found applications in a variety of models relevant to cosmology \cite{cosmology} and astrophysics \cite{astro}. In particular, the fact that Galilean interactions allow for a successful screening of the force mediated by the Galileon on solar system scales, via the so-called Vainshtein mechanism \cite{Vainshtein:1972sx}, implies that these theories can in principle modify gravity in a way that is consistent with the most stringent experimental tests of GR. Nevertheless, the study of spherically symmetric backgrounds in Galileon theories has revealed a number of undesirable features related to low strong coupling scales and superluminal propagation of fluctuations in the Galileon field. This has given rise to a wealth of studies that attempt to generalize the original Galileon model in a way that avoids these unpleasant issues while keeping its virtues. These include covariant completions \cite{covariant}, DBI Galileon theories \cite{dbi}, and multifield generalizations \cite{Deffayet:2010zh,multi-field} (see \cite{Trodden:2011xh,deRham:2012az} for reviews).

One particular model in this class is the bi-Galileon theory proposed in \cite{Padilla:2010de,Padilla:2010tj}. This model is the most general extension of the original Galileon theory to two scalar fields, in that it assumes only Galilean invariance (with second-order equations of motion) and a linear coupling between the Galileons and matter. After developing some elegant algebraic methods, the authors of \cite{Padilla:2010de,Padilla:2010tj} analyzed the behavior of perturbations on a spherically symmetric background with a massive point source, concluding that in this setup it is possible to avoid the issues of strong coupling and superluminality that affected the single-Galileon model. The goal of the present paper is to show that these conclusions need to be revised. In particular, we will show that superluminality cannot be avoided in bi-Galileon theory. We will also show that, even in the extension of this model to an arbitrary number of fields (what we will refer to as multi-Galileon theory), it is not possible to find a set of parameters that makes the theory free of superluminal propagation while avoiding the presence of extremely subluminal fluctuations and the related issues of strong coupling.

When this work was in its final stages Ref.\ \cite{deFromont:2013iwa} appeared. This paper has essentially the same scope as ours, and contains a proof of the presence of superluminal perturbations in the cubic multi-Galileon theory, as well as in the quartic bi-Galileon theory, in full agreement with our results. Our work can therefore be seen as a complement to that reference, since it studies the general multi-Galileon theory including both cubic and quartic interactions. In addition, Ref.\ \cite{deFromont:2013iwa} also studies the behavior of perturbations around a gas of particles with a given radial profile, finding that superluminal fluctuations cannot be avoided. This neatly illustrates that the phenomenon of superluminality is not a special feature of the configuration with a massive point source, but a very generic property of Galileon theories.\footnote{After Ref.\ \cite{deFromont:2013iwa} appeared an erratum has been added to Ref.\ \cite{Padilla:2010tj}. This includes a calculation that supports both the results of \cite{deFromont:2013iwa} and of this paper.}

\section{Multi-Galileon theory}  \label{sec:multiGalileon}

The multi-Galileon theory considered here is perhaps the simplest generalization of the original Galileon model \cite{Nicolis:2008in} to $N$ fields. The theory is invariant under independent Galilean transformations in the fields, $\pi_A\to \pi_A+b_A^{\mu}x_{\mu}+c_A$, with $A=1,\ldots,N$, and assumes a universal linear coupling between the Galileons and the trace of the stress-energy tensor. Then, by means of some trivial field redefinitions, we can always condider a single Galileon, say $\pi_1$, to be directly coupled to matter, so that the mixing between the Galileons and matter is encoded in the Lagrangian
\begin{equation}
\mathcal{L}_{\pi,\mathrm{matter}}=\pi_1 T.
\end{equation}
As it was done in \cite{Nicolis:2008in,Padilla:2010de}, we make the working assumption that the contribution of the Galileons to the stress-energy tensor is negligible compared to the effects coming from $\mathcal{L}_{\pi,\mathrm{matter}}$. In other words, we neglect the gravitational backreaction of the Galileon fields. This allows for a consistent local analysis in which all fields can be taken to propagate in flat spacetime. For this theory to be an interesting and viable modification to GR, it is necessary to take into account the nonlinear interactions of the Galileons with themselves, allowing in principle for a successful Vainshtein screening on solar system scales. These interactions are encoded in the Galileon Lagrangian, given in 4 dimensions by
\begin{equation}
\mathcal{L}_{\pi}=\sum_{n=1}^5 \mathcal{L}_n,
\end{equation}
where \cite{Deffayet:2010zh,Padilla:2010de,Deffayet:2009mn,deFromont:2013iwa}
\begin{equation}
\mathcal{L}_n=\sum_{m_1+\cdots+m_N=n-1}\left( \alpha_{m_1,\ldots,m_N}^1\pi_1+\cdots+\alpha_{m_1,\ldots,m_N}^N\pi_N \right)\mathcal{E}_{m_1,\ldots,m_N},
\end{equation}
and
\begin{equation}
\begin{split}
\mathcal{E}_{m_1,\ldots,m_N}&= (m_1+\cdots+m_N)! \delta^{\mu_1}_{[\alpha_1}\cdots\delta^{\mu_{m_1}}_{\alpha_{m_1}}\cdots\delta^{\nu_1}_{\beta_1}\cdots\delta^{\nu_{m_N}}_{\beta_{m_N}]}\\
&~~~~\times [(\partial_{\mu_1}\partial^{\alpha_1}\pi_1)\cdots(\partial_{\mu_{m_1}}\partial^{\alpha_{m_1}}\pi_1)]\cdots[(\partial_{\nu_1}\partial^{\beta_1}\pi_N)\cdots(\partial_{\nu_{m_N}}\partial^{\beta_{m_N}}\pi_N)].
\end{split}
\end{equation}
Explicitly we have
\begin{equation}
\begin{split}
{\mathcal{E}}_{0,\ldots,0}&=1,\\
{\mathcal{E}}_{1,0,\ldots,0}&=[\Pi_1],\\
{\mathcal{E}}_{1,1,0,\ldots,0}&=[\Pi_1][\Pi_2]-[\Pi_1\Pi_2],\\
{\mathcal{E}}_{1,1,1,0,\ldots,0}&=[\Pi_1][\Pi_2][\Pi_3]-[\Pi_1][\Pi_2\Pi_3]-[\Pi_2][\Pi_1\Pi_3]-[\Pi_3][\Pi_1\Pi_2]+[\Pi_1\Pi_2\Pi_3]+[\Pi_1\Pi_3\Pi_2],\\
{\mathcal{E}}_{1,1,1,1,0,\ldots,0}&=[\Pi_1][\Pi_2][\Pi_3][\Pi_4]-[\Pi_1][\Pi_2][\Pi_3\Pi_4]-[\Pi_1][\Pi_3][\Pi_2\Pi_4]-[\Pi_1][\Pi_4][\Pi_2\Pi_3]\\
&~~~~-[\Pi_2][\Pi_3][\Pi_1\Pi_4]-[\Pi_2][\Pi_4][\Pi_1\Pi_3]-[\Pi_3][\Pi_4][\Pi_1\Pi_2] + [\Pi_1][\Pi_2\Pi_3\Pi_4]\\
&~~~~+[\Pi_1][\Pi_2\Pi_4\Pi_3]+[\Pi_2][\Pi_1\Pi_3\Pi_4]+[\Pi_2][\Pi_1\Pi_4\Pi_3]+[\Pi_3][\Pi_1\Pi_2\Pi_4]+[\Pi_3][\Pi_1\Pi_4\Pi_2]\\
&~~~~+[\Pi_4][\Pi_1\Pi_2\Pi_3]+[\Pi_4][\Pi_1\Pi_3\Pi_2] + [\Pi_1\Pi_2][\Pi_3\Pi_4]+[\Pi_1\Pi_3][\Pi_2\Pi_4]+[\Pi_1\Pi_4][\Pi_2\Pi_3]\\
&~~~~- [\Pi_1\Pi_2\Pi_3\Pi_4]-[\Pi_1\Pi_2\Pi_4\Pi_3]-[\Pi_1\Pi_3\Pi_2\Pi_4]-[\Pi_1\Pi_3\Pi_4\Pi_2]-[\Pi_1\Pi_4\Pi_2\Pi_3]\\
&~~~~-[\Pi_1\Pi_4\Pi_3\Pi_2].\\
\end{split}
\end{equation}
All the other Galilean invariants are obtained by exchanging or identifying different fields. We are using the notation $(\Pi_A)^{\mu}_{~~\nu}\equiv \partial^{\mu}\partial_{\nu}\pi_A$, and $[\Pi]$ denotes the trace of the matrix $\Pi$.

It is actually more convenient not to use the Lagrangian coefficients $\alpha_{m_1,\ldots,m_N}^A$ directly, but to use instead the coefficients appearing in the equations of motion as the parametrization of the theory.\footnote{Here and in the following we will use essentially the same notation as the one introduced in \cite{Padilla:2010de,Padilla:2010tj}.} These are given by\footnote{In the following expressions, $\alpha_{m_1,\ldots,m_N}^A$ is defined as zero whenever one of the indices $m_A$ equals $-1$.}
\begin{equation}
\begin{split}
\frac{\delta}{\delta\pi_A}\int d^4x \mathcal{L}_{\pi} &= \sum_{0\leq m_1+\cdots+m_N\leq 4} (m_A+1)\left(\alpha_{m_1,\ldots,m_N}^A+\sum_{B\neq A}\alpha_{m_1,\ldots,m_B-1,\ldots,m_A+1,\ldots,m_N}^B\right)\mathcal{E}_{m_1,\ldots,m_N}\\
&= \sum_{0\leq m_1+\cdots+m_N\leq 4} a_{m_1,\ldots,m_N}^A \mathcal{E}_{m_1,\ldots,m_N},\\
\end{split}
\end{equation}
and we see that the Galileon coefficients $a_{m_1,\ldots,m_N}^A$ are related to the original Lagrangian coefficients by
\begin{equation} \label{eq:lagrangian_coeffs}
a_{m_1,\ldots,m_N}^A = (m_A+1)\left(\alpha_{m_1,\ldots,m_N}^A+\sum_{B\neq A}\alpha_{m_1,\ldots,m_B-1,\ldots,m_A+1,\ldots,m_N}^B\right).
\end{equation}
These coefficients are not all independent, however, since they satisfy some integrability conditions because of the fact that they are all derived from the same Lagrangian \cite{Padilla:2010de}. The reader can check that, from Eq.\ (\ref{eq:lagrangian_coeffs}), one obtains
\begin{equation} \label{eq:int_conditions}
m_B a^A_{m_1,\ldots,m_A-1,\ldots,m_B,\ldots,m_N}=m_A a^B_{m_1,\ldots,m_A,\ldots,m_B-1,\ldots,m_N}.
\end{equation}
The equations of motion for the Galileon fields can then be written as
\begin{equation} \label{eq:eoms1}
\begin{split}
\sum_{0\leq m_1+\cdots+m_N\leq4}a^1_{m_1,\ldots,m_N}{\mathcal{E}}_{m_1,\ldots,m_N}&=-T,\\
\sum_{0\leq m_1+\cdots+m_N\leq4}a^A_{m_1,\ldots,m_N}{\mathcal{E}}_{m_1,\ldots,m_N}&=0,~~~~~~(A=2,\ldots,N).
\end{split}
\end{equation}
In the following we will set the tadpole coefficients $a_{0,\ldots,0}^A$ to zero. The reason for this is that we are implicitly thinking in field configurations on a self-accelerating de Sitter background; this was the initial motivation of the original Galileon model. The equations of motion for these configurations have precisely the same form as in (\ref{eq:eoms1}), except that the constant terms do not appear because of the background equations \cite{Nicolis:2008in}.

\section{Spherically symmetric solutions and perturbations} \label{sec:perturbations}

\subsection{Static spherically symmetric solutions}

For static, spherically symmetric field configurations we assume $\pi_A=\pi_A(r)$. The Galilean invariants ${\mathcal{E}}_{m_1,\ldots,m_N}$ then reduce to
\begin{equation}
\begin{split}
{\mathcal{E}}_{1,0,\ldots,0}&=\frac{1}{r^2}\frac{d}{dr}(r^2\pi_1'),\\
{\mathcal{E}}_{1,1,0,\ldots,0}&=\frac{2}{r^2}\frac{d}{dr}(r\pi_1'\pi_2'),\\
{\mathcal{E}}_{1,1,1,0,\ldots,0}&=\frac{2}{r^2}\frac{d}{dr}(\pi_1'\pi_2'\pi_3'),\\
{\mathcal{E}}_{1,1,1,1,0,\ldots,0}&=0,
\end{split}
\end{equation}
and all the other terms can be obtained by exchanging or identifying different fields. Note that the prime denotes differentiation with respect to $r$. We will focus on the case of a massive point source, for which $T=-M\delta^3(\mathbf{r})$, with $M$ the mass of the source. The equations of motion (\ref{eq:eoms1}) then become
\begin{equation} \label{eq:eoms2}
\begin{split}
\frac{1}{r^2}\frac{d}{dr}\left(r^3 F^1(y_1,\ldots,y_N)\right)&=M\delta^3(\mathbf{r}),\\
\frac{1}{r^2}\frac{d}{dr}\left(r^3 F^A(y_1,\ldots,y_N)\right)&=0,~~~~~~(A=2,\ldots,N)\\
\end{split}
\end{equation}
where
\begin{equation}
F^A(y_1,\ldots,y_N)\equiv f_1^A+2f_2^A+2f_3^A,
\end{equation}
\begin{equation}
f_n^A\equiv \sum_{m_1+\cdots+m_N=n}a^A_{m_1,\ldots,m_N}y_1^{m_1}\cdots y_N^{m_N},
\end{equation}
and we defined $y_A=\pi_A'/r$. Integrating Eq.\ (\ref{eq:eoms2}) gives the following $N$ algebraic equation for $y_1,\ldots,y_N$:
\begin{equation}
\begin{split}
F^1(y_1,\ldots,y_N)&=\frac{M}{4\pi r^3},\\
F^A(y_1,\ldots,y_N)&=0,~~~~~~(A=2,\ldots,N).
\end{split}
\end{equation}
Notice that both the coefficients $a^A_{m_1,\ldots,m_N}$ and the variables $y_A$ are dimensionful. The scales of the Galileon coefficients are chosen to be such that, on the de Sitter background, the nonlinearities in the fields become important on Hubble scales and, in addition, that it is consistent to neglect the gravitational backreaction of the Galileons. Although this is not relevant for our present purposes, let us simply mention that this choice of scales results in a Vainshtein radius of \cite{Nicolis:2008in,Padilla:2010tj}
\begin{equation} \label{eq:vainshtein_r}
r_V\sim \left(\frac{M}{M_{\mathrm{Pl}}^2H_0^2}\right)^{1/3}.
\end{equation}
With this in mind, we can now perform some trivial rescalings and make the Galileon coefficients $a^A_{m_1,\ldots,m_N}$ dimensionless and (generically) of $O(1)$, and to make the variables $y_A$ dimensionless as well. With a slight abuse of notation we will keep using the same symbols though. The equations of motion then become
\begin{equation} \label{eq:eoms3}
\begin{split}
F^1(y_1,\ldots,y_N)&=\left(\frac{r_V}{r}\right)^3,\\
F^A(y_1,\ldots,y_N)&=0,~~~~~~(A=2,\ldots,N).
\end{split}
\end{equation}
We see that the size of the variables $y_A$ is directly tied to the ratio $r_V/r$ in this notation. Next we define the matrices $\Sigma_n$ with entries
\begin{equation}
(\Sigma_n)_{AB}\equiv \frac{\partial}{\partial y_A}f_n^B.
\end{equation}
With the help of Eq.\ (\ref{eq:int_conditions}) it is easy to show that the matrices $\Sigma_n$ are symmetric. Notice that $\Sigma_2$, $\Sigma_3$, and $\Sigma_4$ depend on the $y_A$, and therefore on $r$, but $\Sigma_1$ is a constant. In fact, the matrix $\Sigma_1$ is the same matrix that appears in the Galileon kinetic Lagrangian:
\begin{equation}
\mathcal{L}_2=-\frac{1}{2}\sum_{A,B}(\Sigma_1)_{AB}\partial^{\mu}\pi_A\partial_{\mu}\pi_B.
\end{equation}
We will therefore require that $\Sigma_1$ be strictly positive definite in order to avoid ghost instabilities. Using these definitions, the Jacobian matrix of the functions $F^A(y_1,\ldots,y_N)$ can be written as
\begin{equation} \label{eq:def_u}
U=\Sigma_1+2\Sigma_2+2\Sigma_3.
\end{equation}
A continuous solution for the $y_A$ exists provided $\mathrm{det}U\neq 0$ for all $r>0$. At large distances from the source, $r\gg r_V$, the $y_A$ are small and the equations of motion are dominated by the linear functions $f_1^A$. Morover, in this regime we have $\mathrm{det}U\simeq \mathrm{det}\Sigma_1>0$, from where we see that $\mathrm{det}U>0$ for all $r>0$ since $\mathrm{det}U$ cannot change sign.

At large distances, Eq.\ (\ref{eq:eoms3}) implies that $y_A\sim (r_V/r)^3$, from where it follows that $\pi_A \sim r_V/r$. This is of course the expected behavior for the regime where nonlinearities are not important. At short distances $r\ll r_V$, on the other hand, the equations of motion are dominated by the cubic functions $f_3^A$, and we generically expect to have $y_A\sim r_V/r$. It follows that $\pi_A \sim r/r_V$ in this regime, which shows that the Galileon fields are indeed screened at distances smaller than the Vainshtein radius.

\subsection{Perturbations of the spherically symmetric background}

The next step in our analysis is to study the behavior of perturbations on the spherically symmetric background discussed above. We let $\pi_A\to \pi_A+\phi_A$, where $\phi_A(t,\mathbf{r})$ is a small fluctuation. To quadratic order the Lagrangian for the fluctuations can be written as
\begin{equation}
{\mathcal{L}}_{\phi}=\frac{1}{2}\partial_t\Phi\cdot K\partial_t\Phi-\frac{1}{2}\partial_r\Phi\cdot U\partial_r\Phi-\frac{1}{2}\partial_{\Omega}\Phi\cdot V\partial_{\Omega}\Phi,
\end{equation}
where $\Phi=(\phi_1,\ldots,\phi_N)$, and $K$, $U$, and $V$ are $N\times N$ matrices. The matrix $U$ is the same matrix that was defined in Eq.\ (\ref{eq:def_u}). In the above equation $\partial_{\Omega}$ denotes the angular part of the gradient operator in spherical coordinates. The matrices $K$ and $V$ can be most easily computed from the equations of motion for the fluctuations, with the result
\begin{equation} \label{eq:matrices_kv}
\begin{split}
K&= \left(1+\frac{1}{3}r\frac{d}{dr}\right)\left( \Sigma_1+3\Sigma_2+6\Sigma_3+6\Sigma_4 \right),\\
V&= \left(1+\frac{1}{2}r\frac{d}{dr}\right)U.
\end{split}
\end{equation}
Notice that, to avoid ghost instabilities in the fluctuations, we must require that the matrix $K$ be positive definite for all $r>0$. It was shown in Ref.\ \cite{Padilla:2010tj} that the quintic Galileon Lagrangian $\mathcal{L}_5$ leads to a very low strong coupling scale on the spherically symmetric background, precluding a consistent analysis in terms of the Galileon fields at short distances from the source. Explicitly, assuming a Vainshtein radius given by Eq.\ (\ref{eq:vainshtein_r}), one can show that the strong coupling scale in the presence of quintic Galileon interactions is
\begin{equation}
\Lambda(r)\sim r^{1/6}\frac{M_{\mathrm{Pl}}^{4/9}H_0^{7/9}}{M^{1/18}}.
\end{equation}
Galileon excitations become strongly coupled when $r\Lambda(r)\sim1$, which for the solar system corresponds to a critical distance $r_c\sim10^5$ km. Even though this distance is much smaller than the Vainshtein radius ($r_V\sim 10^{16}$ km for the Sun), it is clear that the domain of applicability of the theory is quite limited, in the sense that the perturbative description in terms of the fields $\pi_A$ can no longer be trusted below the distance $r_c$. We will therefore adopt the assumption that these quintic interactions are absent, so that $\mathcal{L}_5=0$. This also implies that $\Sigma_4=0$ in Eq.\ (\ref{eq:matrices_kv}).

The linearized equation of motion for the perturbations reads
\begin{equation}
-K\partial_t^2\Phi+\frac{1}{r^2}\partial_r\left(r^2U\partial_r\Phi\right)+V\partial_{\Omega}^2\Phi=0,
\end{equation}
where $\partial_{\Omega}^2$ denotes de angular part of the Laplacian operator. For perturbations of sufficiently small scales, we can approximate these equations as
\begin{equation}
-K\partial_t^2\Phi+U\partial_r^2\Phi+V\partial_{\Omega}^2\Phi=0.
\end{equation}
In Fourier space we have
\begin{equation}
\left[K\omega^2-Up_r^2-Vp_{\Omega}^2\right]\tilde{\Phi}(\omega,p_r,p_{\Omega})=0,
\end{equation}
where $p_r$ and $p_{\Omega}$ are, respectively, the momenta along the radial and orthoradial directions. Parametrizing the momenta as $p_r=p\cos q$, $p_{\Omega}=p\sin q$, we find that the squared sound speeds $c_A^2(q)$ ($A=1,\ldots,N$) are given by the eigenvalues of the matrix
\begin{equation}
M(q)=K^{-1}U\cos^2 q+K^{-1}V\sin^2 q.
\end{equation}
Absence of gradient instabilities requires that $c_A^2(q)\geq0$ for all $r>0$ and for all $q$. This means that the matrix $M(q)$ must have nonnegative eigenvalues for all $r$ and $q$. Absence of superluminal perturbations requires that $c_A^2(q)\leq1$ for all $r$ and $q$. This means that the matrix $M(q)-I$ must have nonpositive eigenvalues for all $r$ and $q$.

One further requirement we would like to impose on the theory is that extremely subluminal modes with $c_A^2\ll1$ be absent. The reason for this is that, as shown in Ref.\ \cite{Nicolis:2008in}, this phenomenon of extreme subluminality is directly associated with a lack of damping in the fluctuations which results in a low strong coupling scale. This is undesirable at the theoretical level, since it significantly limits the domain of validity of the perturbative analysis; outside this domain the Galileon fields $\pi_A$ cannot be reliably identified as the correct low energy degrees of freedom of the theory. Furthermore, there is also the phenomenological issue of the Cerenkov radiation associated with, for instance, the motion of the Earth in the field of the Sun. This has the implication that the Galileon field around the Earth cannot be consistently computed in the static approximation due to the retardation effects related to the slow propagation of the Galileon fluctuations.

In the following subsections we will study the behavior of perturbations in the large- and short-distance limits. We will show that the above requirements cannot all be satisfied simultaneously. This is the main result of this paper.

\subsection{Behavior of perturbations at large distances}

We begin by studying the behavior of fluctuations at large distances from the source, distances much larger than the Vainshtein radius $r_V$. This is the regime where the variables $y_A$ are small, and therefore the equations of motion are dominated by the linear terms $f_1^A$. Following \cite{Padilla:2010tj}, we perform an asymptotic expansion in decreasing powers of $r$ of the form $y_A=y_A^{(0)}+y_A^{(1)}+\ldots$. Let us assume first that neither the linear terms $f_1^A(y_A^{(0)})$ vanish (this is equivalent to assuming that the kinetic terms for the fluctuations do not vanish), nor the quadratic terms $f_2^A(y_A^{(0)})$ vanish. From the equations of motion we then find
\begin{equation}
y_A^{(0)}\propto \frac{1}{r^3},~~~~~~y_A^{(1)}\propto \frac{1}{r^6},~~~~\ldots
\end{equation}
Expanding the matrices $\Sigma_n$ perturbatively in a similar fashion, $\Sigma_n=\Sigma_n^{(0)}+\Sigma_n^{(1)}+\ldots$, we find
\begin{equation}
\Sigma_2^{(0)}\propto \frac{1}{r^3},~~~~~~\Sigma_2^{(1)},\Sigma_3^{(0)}\propto \frac{1}{r^6},~~~~\ldots
\end{equation}
The matrices $K$, $U$, and $V$ are approximately given by
\begin{equation}
\begin{split}
K&\simeq \Sigma_1-3\Sigma_2^{(1)}-6\Sigma_3^{(0)},\\
U&\simeq \Sigma_1+2\Sigma_2^{(0)}+2\left(\Sigma_2^{(1)}+\Sigma_3^{(0)}\right),\\
V&\simeq \Sigma_1-\Sigma_2^{(0)}-4\left(\Sigma_2^{(1)}+\Sigma_3^{(0)}\right),
\end{split}
\end{equation}
where terms of order $(r_V/r)^9$ were neglected. The matrix $M(q)$ is given by
\begin{equation}
\begin{split}
M(q)&\simeq I+(3\cos^2q-1)\Sigma_1^{-1}\Sigma_2^{(0)}+(6\cos^2q-1)\Sigma_1^{-1}\Sigma_2^{(1)}+(6\cos^2q+2)\Sigma_1^{-1}\Sigma_3^{(0)}.
\end{split}
\end{equation}
We see that the matrix $M(q)-I\simeq(3\cos^2q-1)\Sigma_1^{-1}\Sigma_2^{(0)}$ changes sign at $\cos^2q=1/3$, implying that perturbations along some directions will be superluminal. We can avoid this by choosing coefficients such that $\Sigma_2^{(0)}=0$. This in turn implies that the quadratic terms $f_2^A(y_A^{(0)})$ of the equations of motion vanish (see Appendix \ref{sec:app1}), and so one must repeat the analysis starting with this assumption.

Assume that the linear terms $f_1^A(y_A^{(0)})$ and the cubic terms $f_3^A(y_A^{(0)})$ do not vanish, but that the quadratic terms $f_2^A(y_A^{(0)})$ as well as the matrix $\Sigma_2^{(0)}$ do vanish. Then we find
\begin{equation}
y_A^{(0)}\propto \frac{1}{r^3},~~~~~~y_A^{(1)}\propto \frac{1}{r^9},~~~~\ldots
\end{equation}
\begin{equation}
\Sigma_3^{(0)}\propto \frac{1}{r^6},~~~~~~\Sigma_2^{(1)}\propto \frac{1}{r^9},~~~~\ldots
\end{equation}
The matrices $K$, $U$, and $V$ are approximately given by
\begin{equation}
\begin{split}
K&\simeq \Sigma_1-6\Sigma_3^{(0)}-6\Sigma_2^{(1)},\\
U&\simeq \Sigma_1+2\Sigma_3^{(0)}+2\Sigma_2^{(1)},\\
V&\simeq \Sigma_1-4\Sigma_3^{(0)}-7\Sigma_2^{(1)},
\end{split}
\end{equation}
where terms of order $(r_V/r)^{12}$ were neglected. The matrix $M(q)$ is given by
\begin{equation}
\begin{split}
M(q)&\simeq I+(6\cos^2q+2)\Sigma_1^{-1}\Sigma_3^{(0)}+(9\cos^2q-1)\Sigma_1^{-1}\Sigma_2^{(1)}.
\end{split}
\end{equation}
It is clear that, for the matrix $M(q)-I$ to have nonpositive eigenvalues, we need that the matrix $\Sigma_3^{(0)}$ be negative semidefinite. This requirement would make perturbations (slightly) subluminal at large distances, and in addition to the condition that $\Sigma_1$ be positive definite, the theory would be free of instabilities in this regime.

\subsection{Behavior of perturbations at short distances}

We repeat the analysis of the previous subsection, this time in the region of short distances from the source, $r\ll r_V$. Again we perform an asymptotic expansion, $y_A=y_A^{(0)}+y_A^{(1)}+\ldots$, this time in increasing powers of $r$. We can then solve the equations of motion order by order, finding that
\begin{equation}
y_A^{(0)}\propto \frac{1}{r},~~~~~~y_A^{(1)}\propto1,~~~~\ldots
\end{equation}
Note that this assumes that the cubic terms $f_3^A(y_A^{(0)})$, which dominate the equations of motion at short distances, as well as the quadratic terms $f_2^A(y_A^{(0)})$, do not vanish. Expanding the matrices $\Sigma_n$ perturbatively, i.e.\ $\Sigma_n=\Sigma_n^{(0)}+\Sigma_n^{(1)}+\ldots$, we find
\begin{equation}
\Sigma_3^{(0)}\propto \frac{1}{r^2},~~~~~~~~\Sigma_3^{(1)},\Sigma_2^{(0)}\propto \frac{1}{r},~~~~\ldots
\end{equation}
The matrices $K$, $U$, and $V$ are approximately given by
\begin{equation} \label{eq:pert_matrices2}
\begin{split}
K&\simeq 2\Sigma_3^{(0)}+2\left(2\Sigma_3^{(1)}+\Sigma_2^{(0)}\right),\\
U&\simeq 2\Sigma_3^{(0)}+2\left(\Sigma_3^{(1)}+\Sigma_2^{(0)}\right),\\
V&\simeq \Sigma_3^{(1)}+\Sigma_2^{(0)},
\end{split}
\end{equation}
where terms of $O(1)$ were neglected. It is clear that $K^{-1}V=O(r/r_V)$ at small distances, implying that there exist fluctuations in the orthoradial direction that are extremely subluminal when $r\ll r_V$ (this is assuming the matrix $K^{-1}V$ has only nonnegative eigenvalues; otherwise there is an instability). The way to avoid this would be to choose the parameters of the theory so that $\Sigma_3^{(0)}=0$. From the results of Appendix \ref{sec:app1}, this is not consistent with the assumption $f_3^A(y_A^{(0)})\neq 0$, so we must repeat the analysis starting from the assumption that both $\Sigma_3^{(0)}=0$ and $f_3^A(y_A^{(0)})=0$ at $r\ll r_V$.\footnote{In principle it could be that $f_3^A(y_A^{(0)})=0$ but $\Sigma_3^{(0)}\neq 0$. It is easy to see, however, that in that case the matrix $K$ would go as $1/r^{3/2}$ at short distances, whereas the matrices $U$ and $V$ would go as $1/r^3$, implying the existence of squared sound speeds that are very large and positive or very large and negative, neither of which is desirable.} The equations of motion then imply that the $y_A$ do not go as $1/r$ at short distances. Instead, assuming that the quadratic terms $f_2^A$ do not vanish\footnote{In Appendix \ref{sec:app5} we show that, for the case of two Galileons, the condition that $\Sigma_2^{(0)}=0$ at large distances implies that the functions $f_2^A$ must vanish in the short-distance limit, implying that there can be no Vainshtein screening of the Galileons for $r\ll r_V$. The following special case therefore applies to multi-Galileon theory with $N\geq3$.} (otherwise we would be left with the linear terms only, and so the Vainshtein mechanism would not work at all), we find\footnote{Notice that to derive that $y_A^{(1)}\propto 1$ we used the fact that $(f_3^A)^{(1)}=0$, as follows from the results of Appendix \ref{sec:app1}.}
\begin{equation}
y_A^{(0)}\propto \frac{1}{r^{3/2}},~~~~~~y_A^{(1)}\propto 1,~~~~\ldots
\end{equation}
This gives
\begin{equation}
\Sigma_3^{(1)},\Sigma_2^{(0)}\propto \frac{1}{r^{3/2}},~~~~\ldots
\end{equation}
and the matrices $K$, $U$, and $V$ are approximately given by
\begin{equation} \label{eq:pert_matrices1}
\begin{split}
K&\simeq 3\Sigma_3^{(1)}+\frac{3}{2}\Sigma_2^{(0)},\\
U&\simeq 2\left(\Sigma_3^{(1)}+\Sigma_2^{(0)}\right),\\
V&\simeq \frac{1}{2}\left(\Sigma_3^{(1)}+\Sigma_2^{(0)}\right),
\end{split}
\end{equation}
where terms of $O(1)$ were neglected. The matrix $M(q)$ is then given by
\begin{equation}
M\simeq \frac{2}{3}\cos^2 q \left(\Sigma_3^{(1)}+\frac{1}{2}\Sigma_2^{(0)}\right)^{-1}\left(\Sigma_3^{(1)}+\Sigma_2^{(0)}\right)+\frac{1}{6}\sin^2 q \left(\Sigma_3^{(1)}+\frac{1}{2}\Sigma_2^{(0)}\right)^{-1}\left(\Sigma_3^{(1)}+\Sigma_2^{(0)}\right).
\end{equation}
Consider the matrix
\begin{equation}
A\equiv \left(\Sigma_3^{(1)}+\frac{1}{2}\Sigma_2^{(0)}\right)^{-1}\left(\Sigma_3^{(1)}+\Sigma_2^{(0)}\right) = 2I-\left(\Sigma_3^{(1)}+\frac{1}{2}\Sigma_2^{(0)}\right)^{-1}\Sigma_3^{(1)}.
\end{equation}
From the results of Appendix \ref{sec:app2}, we have that $\mathrm{det}\Sigma_3^{(1)}=0$, from where it follows that the matrix $A$ has at least one eigenvalue equal to $2$. This implies that the matrix $M$ has an eigenvalue equal to $4/3$ when $\sin q=0$, corresponding to a superluminal mode along the radial direction.

We conclude that by tuning the Galileon coefficients in such a way as to avoid the extremely subluminal perturbations that generically appear near the source, we end up in turn with a superluminal mode. It seems preferable, then, to go back to the assumption that $f_3^A(y_A^{(0)})\neq 0$, which implies that the matrix $\Sigma_3^{(0)}$ does not vanish identically, and simply accept the presence of an extremely subluminal mode (forgetting about the undesirable phenomenological and theoretical consequences of this). We still need to make sure, of course, that there are neither unstable nor superluminal modes. It is clear, from Eq.\ (\ref{eq:pert_matrices2}), that we can avoid a gradient instablity and guarantee the existence of continuous solutions by choosing parameters such that $\Sigma_3^{(0)}$ is positive semidefinite. Notice that, by the results of Appendix \ref{sec:app3}, we cannot have $\Sigma_3^{(0)}$ {\it strictly} positive definite for $r\ll r_V$, since the absence of superluminality at large distances requires $\Sigma_3^{(0)}$ to be negative semidefinite for $r\gg r_V$. The only possible loophole would be that, by a very special choice of the Galileon parameters, $\Sigma_3^{(0)}$ results to be singular positive semidefinite (but nonvanishing) for $r\ll r_V$, and that it results to be singular negative semidefinite for $r\gg r_V$. We have no proof that, for an arbitrary number of Galileon fields, this special case also suffers from instablities or superluminality (in Appendix \ref{sec:app4} it is shown that this loophole leads to a contradiction in the case of two Galileons). The main conclusion remains however: extremely subluminal perturbations will be present at distances close to the source.

\section{Final remarks} \label{sec:summary}

Let us briefly summarize our results. We have studied the behavior of small fluctuations on static, spherically symmetric backgrounds in multi-Galileon theory. By imposing the requirements of stability, successful Vainshtein screening, absence of superluminality, and absence of extremely subluminal fluctuations, we have derived constraints in the large- and short-distance regimes. We have shown that these constraints cannot all be satisfied simultaneously in multi-Galileon theory and, in particular, that superluminal perturbations will always be present in bi-Galileon theory. It is an interesting open problem whether superluminality can in principle be avoided in multi-Galileon theory with $N\geq3$, thanks to the loophole we identified at the end of Sec.\ \ref{sec:perturbations}. It seems that neither our results of Appendix \ref{sec:app4} nor the proof given in \cite{deFromont:2013iwa} can be easily generalized to more than two Galileons. Perhaps the geometrical picture we developed in Appendix \ref{sec:app3} will prove useful to solve this question.

One may argue that the issues related to the retardation effects due to the extremely subluminal fluctuations are not serious, since after all the Galileon fields are screened at short distances and the corresponding Cerenkov radiation may not present a problem from the experimental point of view \cite{Padilla:2010tj}. Of course, there is still the question of whether such a model can be consistently applied in the static approximation to the solar system, but this is not directly related to the consistency of the model at the theoretical level. The problem, however, is that the slow propagation of these fluctuations implies a lack of enhancement of the kinetic Lagrangian relative to the interactions, with the consequence of a very low strong-interaction scale for the fluctuations \cite{Nicolis:2008in}. Perhaps more serious is the presence of superluminal perturbations, which were shown to be unavoidable in the bi-Galileon theory, and that will generically be present also in multi-Galileon theory. On the one hand, superluminality is a sign that the theory cannot be regarded as the effective low-energy description of a microscopic Lorentz-invariant theory \cite{superluminality}. However, this by itself does not mean that the theory is inconsistent, of course, unless one shows that closed timelike curves exist in the regime where the effective field theory is applicable. Interestingly, it has been conjectured that this does not happen in Galileon theory \cite{Burrage:2011cr}, although it seems that further investigation is needed before we can reach a full understanding of the physical origin of superluminality in this class of theories.

\acknowledgments The author would like to thank Claudia de Rham and Lavinia Heisenberg for useful comments, and especially Alberto Nicolis for a number of instructive and helpful conversations throughout this work.

\appendix
\section{} \label{sec:app1}

In this section we show that the condition $\Sigma_n^{(0)}=0$ implies that $f_n^A(y_B^{(0)})=0$ for all $A=1,\ldots,N$. From the definition
\begin{equation}
f_n^A=\sum_{m_1+\cdots+m_N=n}a^A_{m_1,\ldots,m_N}y_1^{m_1}\cdots y_N^{m_N},
\end{equation}
we have
\begin{equation}
y_B\frac{\partial}{\partial y_B}f_n^A=\sum_{m_1+\cdots+m_N=n}m_Ba^A_{m_1,\ldots,m_N}y_1^{m_1}\cdots y_N^{m_N}.
\end{equation}
Therefore
\begin{equation}
\begin{split}
\sum_{B=1}^N y_B \frac{\partial}{\partial y_B}f_n^A &= \sum_{m_1+\cdots+m_N=n}a^A_{m_1,\ldots,m_N}y_1^{m_1}\cdots y_N^{m_N}(m_1+\cdots+m_N)\\
&=n\sum_{m_1+\cdots+m_N=n}a^A_{m_1,\ldots,m_N} y_1^{m_1}\cdots y_N^{m_N}\\
&=n f_n^A.
\end{split}
\end{equation}
It follows that the condition
\begin{equation}
(\Sigma_n^{(0)})_{BA}=\frac{\partial}{\partial y_B}f_n^{A}\Bigg|_{y=y^{(0)}}=0
\end{equation}
implies that $f_n^A(y_B^{(0)})=0$ for all $A=1,\ldots,N$. Conversely, the assumption that $f_n^A(y_B^{(0)})\neq 0$ for some $A$ implies that the matrix $\Sigma_n^{(0)}$ cannot vanish identically.

Incidentally, notice that the condition $\Sigma_n^{(0)}=0$ also implies that the next-to-leading functions $(f_n^A)^{(1)}$ also vanish. Indeed,
\begin{equation}
(f_n^A)^{(1)}=\sum_{B=1}^N y_B^{(1)}\frac{\partial}{\partial y_B}f_n^{A}\Bigg|_{y=y^{(0)}}=\sum_{B=1}^N y_B^{(1)}(\Sigma_n^{(0)})_{BA}=0.
\end{equation}

\section{} \label{sec:app2}

In this section we prove that $\det\Sigma_n^{(1)}=0$ if $\Sigma_n^{(0)}=0$. Recall that we are working with an asymptotic expansion (either at large or short distances from the source) of the form $y_A=y_A^{(0)}+y_A^{(1)}+\ldots$. We can assume that all the $y_A^{(0)}$ are proportional to each other, $y_A^{(0)}=\alpha_A z^{(0)}$, and likewise for the $y_A^{(1)}$, that is $y_A^{(1)}=\beta_A z^{(1)}$. The $\alpha_A$ cannot be all equal to zero, and in particular we can set one of them equal to 1. The same holds for the $\beta_A$. Recall the definition
\begin{equation}
(\Sigma_n)_{BA}=\frac{\partial}{\partial y_B}f_n^A= \sum_{m_1+\cdots+m_N=n} m_B a^A_{m_1,\ldots,m_N}y_1^{m_1}\cdots y_B^{m_B-1}\cdots y_N^{m_N}.
\end{equation}
We are assuming that $\Sigma_n^{(0)}$ vanishes, and therefore
\begin{equation} \label{eq:app1}
(\Sigma_n^{(0)})_{BA}=(z^{(0)})^{n-1} \sum_{m_1+\cdots+m_N=n}m_B a^A_{m_1,\ldots,m_N}\alpha_1^{m_1}\cdots\alpha_B^{m_B-1}\cdots\alpha_N^{m_N}=0.
\end{equation}
The matrix $\Sigma_n^{(1)}$ is given by
\begin{equation}
\begin{split}
(\Sigma_n^{(1)})_{BA}&=\sum_{m_1+\cdots+m_N=n}m_B a^A_{m_1,\ldots,m_N} \Big[m_1y_1^{(1)}y_1^{(0)m_1-1}\cdots y_B^{(0)m_B-1}\cdots y_N^{(0)m_N} +\cdots\\
&~~~~ + (m_B-1)y_B^{(1)}y_1^{(0)m_1}\cdots y_B^{(0)m_B-2}\cdots y_N^{(0)m_N} +\cdots\\
&~~~~ + m_Ny_N^{(1)}y_1^{(0)m_1}\cdots y_B^{(0)m_B-1}\cdots y_N^{(0)m_N-1}\Big]\\
&=(z^{(0)})^{n-2}z^{(1)}\sum_{m_1+\cdots+m_N=n}m_B a^A_{m_1,\ldots,m_N} \Big[m_1\beta_1\alpha_1^{m_1-1}\cdots \alpha_B^{m_B-1}\cdots \alpha_N^{m_N} +\cdots\\
&~~~~ + (m_B-1)\beta_B\alpha_1^{m_1}\cdots \alpha_B^{m_B-2}\cdots \alpha_N^{m_N} +\cdots+ m_N\beta_N\alpha_1^{m_1}\cdots \alpha_B^{m_B-1}\cdots \alpha_N^{m_N-1}\Big].\\
\end{split}
\end{equation}
Consider the linear combination
\begin{equation}
\begin{split}
\sum_{B=1}^N \alpha_B(\Sigma_n^{(1)})_{BA} &=(z^{(0)})^{n-2}z^{(1)}\sum_{B=1}^N\sum_{m_1+\cdots+m_N=n}m_B a^A_{m_1,\ldots,m_N} \Big[m_1\beta_1\alpha_1^{m_1-1}\cdots\alpha_N^{m_N} +\cdots\\
&~~~~+ m_N\beta_N\alpha_1^{m_1}\cdots\alpha_N^{m_N-1}-\beta_B\alpha_1^{m_1}\cdots \alpha_B^{m_B-1}\cdots \alpha_N^{m_N}\Big]\\
&=n(z^{(0)})^{n-2}z^{(1)}\sum_{m_1+\cdots+m_N=n}a^A_{m_1,\ldots,m_N}\Big[m_1\beta_1\alpha_1^{m_1-1}\cdots\alpha_N^{m_N} +\cdots\\
&~~~~ + m_N\beta_N\alpha_1^{m_1}\cdots\alpha_N^{m_N-1}\Big]\\
&~~~~ -(z^{(0)})^{n-2}z^{(1)}\sum_{B=1}^N \beta_B \sum_{m_1+\cdots+m_N=n}m_B a^A_{m_1,\ldots,m_N}\alpha_1^{m_1}\cdots\alpha_B^{m_B-1}\cdots\alpha_N^{m_N}\\
&=0,
\end{split}
\end{equation}
from Eq.\ (\ref{eq:app1}). This shows that the rows of the matrix $\Sigma_n^{(1)}$ are not linearly independent, and therefore $\det \Sigma_n^{(1)}=0$.

\section{} \label{sec:app3}

In this section we prove that the matrix $\Sigma_3$ (thought as a function of the $N$ variables $y_A$) cannot be positive definite at one point, and be negative semidefinite at some other point. The key observation is that the $f_3^A$ can all be derived from a single function $L_3$. To see this, start from
\begin{equation}
\frac{\partial L_3}{\partial y_1}=f_3^1=\sum_{m_1+\cdots+m_N=3}a^1_{m_1,\ldots,m_N}y_1^{m_1}\cdots y_N^{m_N},
\end{equation}
and integrate to find
\begin{equation}
\begin{split}
L_3&=\sum_{m_1+\cdots+m_N=3}\frac{a^1_{m_1,\ldots,m_N}}{m_1+1}y_1^{m_1+1}\cdots y_N^{m_N}+g_1(y_A\neq y_1)\\
&=\sum_{\substack{{m_1+\cdots+m_N=4}\\{m_1\neq0}}}\frac{a^1_{m_1-1,\ldots,m_N}}{m_1}y_1^{m_1}\cdots y_N^{m_N}+g_1(y_A\neq y_1).\\
\end{split}
\end{equation}
Here $g_1$ is some function that depends on all the $y_A$ except $y_1$. Differentiating with respect to $y_2$ we obtain
\begin{equation}
\begin{split}
\frac{\partial L_3}{\partial y_2}&=\sum_{\substack{{m_1+\cdots+m_N=4}\\{m_1\neq0}}}\frac{a^1_{m_1-1,\ldots,m_N}}{m_1}m_2y_1^{m_1}y_2^{m_2-1}\cdots y_N^{m_N}+\frac{\partial g_1}{\partial y_2}\\
&=\sum_{\substack{{m_1+\cdots+m_N=3}\\{m_1\neq0}}}{a^2_{m_1,\ldots,m_N}}y_1^{m_1}\cdots y_N^{m_N}+\frac{\partial g_1}{\partial y_2},\\
\end{split}
\end{equation}
where we used Eq.\ (\ref{eq:int_conditions}). Equating this with $f_3^2$, we get
\begin{equation}
\frac{\partial g_1}{\partial y_2}=\sum_{\substack{{m_1+\cdots+m_N=3}\\{m_1=0}}}{a^2_{m_1,\ldots,m_N}}y_1^{m_1}\cdots y_N^{m_N},
\end{equation}
which is independent of $y_1$. Integrating again gives
\begin{equation}
\begin{split}
g_1&=\sum_{\substack{{m_1+\cdots+m_N=3}\\{m_1=0}}}\frac{a^2_{m_1,\ldots,m_N}}{m_2+1}y_1^{m_1}y_2^{m_2+1}\cdots y_N^{m_N}+g_2(y_A\neq y_1,y_2)\\
&=\sum_{\substack{{m_1+\cdots+m_N=4}\\{m_1=0,m_2\neq0}}}\frac{a^2_{m_1,m_2-1\ldots,m_N}}{m_2}y_1^{m_1}\cdots y_N^{m_N}+g_2(y_A\neq y_1,y_2),\\
\end{split}
\end{equation}
where $g_2$ is a function that depends on all the $y_A$ except $y_1$ and $y_2$. Repeating this process $N$ times finally gives the function $L_3$ (up to an irrelevant integration constant):
\begin{equation}
\begin{split}
L_3&=\sum_{\substack{{m_1+\cdots+m_N=4}\\{m_1\neq0}}}\frac{a^1_{m_1-1,\ldots,m_N}}{m_1}y_1^{m_1}\cdots y_N^{m_N}+\sum_{\substack{{m_1+\cdots+m_N=4}\\{m_1=0,m_2\neq0}}}\frac{a^2_{m_1,m_2-1\ldots,m_N}}{m_2}y_1^{m_1}\cdots y_N^{m_N}\\
&~~~~+\cdots+\sum_{\substack{{m_1+\cdots+m_N=4}\\{m_1=0,\ldots,m_{N-1}=0,m_N\neq0}}}\frac{a^N_{m_1,\ldots,m_N-1}}{m_N}y_1^{m_1}\cdots y_N^{m_N}.\\
\end{split}
\end{equation}
From Eq.\ (\ref{eq:int_conditions}) we see that the coefficients appearing in these sums are all independent, and so the complete sum can be written as
\begin{equation}
L_3=\sum_{m_1+\cdots+m_N=4}A_{m_1,\ldots,m_N}y_1^{m_1}\cdots y_N^{m_N},
\end{equation}
where the $A_{m_1,\ldots,m_N}$ are all independent parameters. Thus, the function $L_3$ is a general $N$-ary quartic form (a homogeneous polynomial of degree 4 in $N$ variables). The entries of the matrix $\Sigma_3$ can then be written as
\begin{equation}
(\Sigma_3)_{AB}=\frac{\partial}{\partial{y_A}}f_3^{B}=\frac{\partial^2}{\partial y_A \partial y_B}L_3,
\end{equation}
that is, $\Sigma_3$ is given by the Hessian matrix of $L_3$. Thus, our task is to prove that the Hessian matrix of a general quartic form cannot be positive definite at some point and negative semidefinite at some other point.

We first show that this is true for $N=2$ (the case $N=1$ is trivially true). We have to show that the Hessian matrix of a general binary quartic form,
\begin{equation}
q(x,y)=ax^4+bx^3y+cx^2y^2+dxy^3+ey^4,
\end{equation}
cannot be positive definite at one point, say $(x_1,y_1)$, and negative semidefinite at some other point, say $(x_2,y_2)$. The proof of this is greatly simplified by the fact that any quartic form can be reduced to its canonical form (see for example \cite{Salmon}),
\begin{equation}
Q(X,Y)=rX^4+6mX^2Y^2+sY^4,
\end{equation}
by means of a nonsingular linear transformation $(x,y)\mapsto (X,Y)$. The Hessian matrix of $Q(X,Y)$ is given by
\begin{equation}
H_Q(X,Y)=12\left( \begin{array}{cc}
rX^2+mY^2 & 2mXY \\
2mXY & sY^2+mX^2 \end{array} \right)
\end{equation}
Assume that there is a point $(X_1,Y_1)\neq(0,0)$ where $H_Q$ is positive definite.\footnote{The condition $(X_1,Y_1)\neq(0,0)$ comes from the requirement that the background Galileon fields do not both vanish.} Then its eigenvalues must be strictly positive, or equivalently, its trace and determinant must be positive:
\begin{equation} \label{eq:tr1}
(rX_1^2+mY_1^2)+(sY_1^2+mX_1^2)>0,
\end{equation}
\begin{equation} \label{eq:det1}
(rX_1^2+mY_1^2)(sY_1^2+mX_1^2)-4m^2X_1^2Y_1^2>0.
\end{equation} 
In particular, these imply that
\begin{equation} \label{eq:cond1}
(rX_1^2+mY_1^2)>0,~~~~~~~~(sY_1^2+mX_1^2)>0.
\end{equation}
Assume also that there is another point $(X_2,Y_2)\neq(0,0)$ where $H_Q$ is negative semidefinite. Then its eigenvalues must be nonpositive, or equivalently, its trace must be nonpositive and its determinant must be nonnegative:
\begin{equation} \label{eq:tr2}
(rX_2^2+mY_2^2)+(sY_2^2+mX_2^2)\leq0,
\end{equation}
\begin{equation} \label{eq:det2}
(rX_2^2+mY_2^2)(sY_2^2+mX_2^2)-4m^2X_2^2Y_2^2\geq0.
\end{equation}
In particular, these imply that
\begin{equation} \label{eq:cond2}
(rX_2^2+mY_2^2)\leq0,~~~~~~~~(sY_2^2+mX_2^2)\leq0.
\end{equation}
It suffices to consider the cases where $(r=1,s=1)$, $(r=-1,s=1)$, and $(r=0,s=1)$ (we leave to the reader to check that in the cases where both $r$ and $s$ are zero, or when $m=0$, one easily arrives at a contradiction).

If $r=1$, $s=1$, then Eqs.\ (\ref{eq:tr1}) and (\ref{eq:tr2}) give
\begin{equation}
(1+m)(X_1^2+Y_1^2)>0,~~~~~~~~(1+m)(X_2^2+Y_2^2)\leq0,
\end{equation}
which is a contradiction.

If $r=-1$, $s=1$, then Eqs.\ (\ref{eq:cond1}) and (\ref{eq:cond2}) give
\begin{equation}
-X_1^2+mY_1^2>0~~~~\Rightarrow~~~~m>\frac{X_1^2}{Y_1^2}\geq0,
\end{equation}
\begin{equation}
Y_2^2+mX_2^2\leq0~~~~\Rightarrow~~~~m\leq -\frac{Y_2^2}{X_2^2}\leq0,
\end{equation}
and we arrive at a contradiction.

If $r=0$, $s=1$, then Eqs.\ (\ref{eq:cond1}) and (\ref{eq:cond2}) give
\begin{equation}
mY_1^2>0,~~~~~~~~ mY_2^2\leq0,
\end{equation}
which implies $m>0$, $Y_2=0$. But if that is the case then Eq.\ (\ref{eq:tr2}) gives
\begin{equation}
mX_2^2\leq0,
\end{equation}
and we arrive again at a contradiction, since $X_2$ and $Y_2$ cannot be both zero.

We conclude that the Hessian $H_Q$ cannot be positive definite at one point and negative semidefinite at some other point. Since positive (or negative) definiteness of the Hessian is a statement about the local convexity (or concavity) of a function, and since convexity is preserved by linear transformations, we conclude that the Hessian of the general quartic form $q(x,y)$ also has this property.

Finally we prove the general case of $N$ variables using induction, by exploiting the relation between the positive definiteness of the Hessian and convexity. Assume that a general $N$-ary quartic form cannot be concave at one point if it is (strictly) convex at some other point. But assume that there is an $(N+1)$-ary quartic form $Q$ that is (strictly) convex at a point $x'=(x_1',\ldots,x_{N+1}')$, and concave at a point $x''=(x_1'',\ldots,x_{N+1}'')$. Consider a hyperplane containing the points $x'$, $x''$ and the origin (such a hyperplane always exists for $N\geq2$; this is why the case of one Galileon does not work as a base case for the inductive argument). Write the equation of this hyperplane by solving for one variable, say $x_A$, in terms of the others (this equation is homogeneous, since the hyperplane contains the origin), and consider the $N$-ary quartic form $Q'$ obtained by constraining $Q$ to this hyperplane. By assumption $Q'$ cannot be strictly convex at $x'$ and concave at $x''$. But then $Q$ cannot satisfy this property either. The reason for this is that a function $f$ is convex (concave) in a region if and only if the function obtained from constraining $f$ to any line contained in that region is convex (concave) also. In particular, for a function $f$ of 3 or more variables, the function obtained from constraining $f$ to any plane (contained in the region where $f$ is convex) must be convex as well. We conclude that the $(N+1)$-ary quartic form $Q$ cannot be strictly convex at $x'$ and concave at $x''$. This completes the inductive argument.

\section{} \label{sec:app5}

In this section we show that, in bi-Galileon theory, the special case in which the matrix $\Sigma_3^{(0)}$ vanishes in the short-distance limit implies the absence of a successful Vainshtein screening of the Galileons. We will use the following notation for the Galileon coefficients: $a_{m_1,m_2}^1\equiv a_{m_1,m_2}$ and $a_{m_1,m_2}^2\equiv b_{m_1,m_2}$. Notice that, with the exception of $b_{01}$, $b_{02}$, and $b_{03}$, all the coefficients $b_{m_1,m_2}$ can be expressed in terms of the coefficients $a_{m_1,m_2}$ by using Eq.\ (\ref{eq:int_conditions}).

Recall that we are working with asymptotic expansions at short distances, $y_A=y_A^{(0)}+y_A^{(1)}+\ldots$, and at large distances, $y_A=\bar{y}_A^{(0)}+\bar{y}_A^{(1)}+\ldots$, where $A=1,2$ in bi-Galileon theory (here and in the following, the bar labels quantities evaluated in the large-distance limit, to avoid confusion with the short-distance limit). Recall also from Sec.\ \ref{sec:perturbations} that we imposed the requirement that $\bar{\Sigma}_2^{(0)}\equiv\Sigma_2(\bar{y}_1^{(0)},\bar{y}_2^{(0)})=0$ in order to avoid superluminal propagation at large distances. In this regime the equations of motion are dominated by the linear functions $f_1^A(\bar{y}_1^{(0)},\bar{y}_2^{(0)})$, and the solutions are easily found to be
\begin{equation} \label{eq:app4_linear_sols}
\bar{y}_1^{(0)}=\frac{b_{01}}{(a_{10}b_{01}-a_{01}^2)}\left(\frac{r_V}{r}\right)^3,~~~~~~~~\bar{y}_2^{(0)}=\frac{-a_{01}}{(a_{10}b_{01}-a_{01}^2)}\left(\frac{r_V}{r}\right)^3.
\end{equation}
The matrix $\bar{\Sigma}_2^{(0)}$ is given by
\begin{equation}
\bar{\Sigma}_2^{(0)}=\left( \begin{array}{cc} 2a_{20}\beta+a_{11} & a_{11}\beta+2a_{02} \\ 
a_{11}\beta+2a_{02} & 2a_{02}\beta+2b_{02} \end{array} \right)\bar{y}_2^{(0)},
\end{equation}
where we defined $\beta\equiv \bar{y}_1^{(0)}/\bar{y}_2^{(0)}=-b_{01}/a_{01}$ (we will come back to the case where $\bar{y}_2^{(0)}=0$ later). The condition $\bar{\Sigma}_2^{(0)}=0$ then implies that the cubic Galileon coefficients can be written in terms of $a_{20}$ and $\beta$ as follows:
\begin{equation} \label{eq:app5_1}
a_{11}=-2a_{20}\beta,~~~~~~a_{02}=a_{20}\beta^2,~~~~~~b_{02}=-a_{20}\beta^3.
\end{equation}

We are also assuming that, at large distances, the matrix $\Sigma_3^{(0)}\equiv\Sigma_3(y_1^{(0)},y_2^{(0)})$ vanishes, as required to avoid extremely subluminal fluctuations. From the results of Appendix \ref{sec:app1} we know that the cubic functions $f_3^A(y_1^{(0)},y_2^{(0)})$ also vanish, and so the equations of motion in the short-distance limit are dominated by the quadratic functions $f_2^A(y_1^{(0)},y_2^{(0)})$. Explicitly we have (see Eq.\ (\ref{eq:eoms3}))
\begin{equation} \label{eq:app5_2}
\begin{split}
2f_2^1(y_1^{(0)},y_2^{(0)})&= 2\left(a_{20}\alpha^2+a11\alpha+a_{02}\right)y_2^{(0)2}=\left(\frac{r_V}{r}\right)^3,\\
2f_2^2(y_1^{(0)},y_2^{(0)})&= 2\left(\frac{a_{11}}{2}\alpha^2+2a_{02}\alpha+b_{02}\right)y_2^{(0)2}=0,
\end{split}
\end{equation}
and we defined $\alpha\equiv y_1^{(0)}/y_2^{(0)}$ (we assume for now that $y_2^{(0)}\neq0$; we will come back to the case where ${y}_2^{(0)}=0$ later). However, if we use relations (\ref{eq:app5_1}) we find that
\begin{equation}
\begin{split}
f_2^1(y_1^{(0)},y_2^{(0)})&= a_{20}\left(\alpha-\beta\right)^2y_2^{(0)2},\\
f_2^2(y_1^{(0)},y_2^{(0)})&= -a_{20}\beta\left(\alpha-\beta\right)^2y_2^{(0)2}.
\end{split}
\end{equation}
Since $\beta\neq0$ (from the requirement of having a positive definite kinetic Lagrangian for the perturbations), the second equation of motion in (\ref{eq:app5_2}) implies that either $a_{20}=0$ (in which case all the cubic Galileon coefficients vanish) or $\alpha=\beta$. In either case we will have that $f_2^1(y_1^{(0)},y_2^{(0)})=0$ identically. This implies that the equations of motion will not be dominated by the quadratic functions $f_2^A(y_1^{(0)},y_2^{(0)})$, but instead they will be dominated by the linear functions $f_1^A(y_1^{(0)},y_2^{(0)})$, meaning that there will be no Vainshtein screening of the Galileons for $r\ll r_V$.

The same conclusions arise in the cases where either ${y}_2^{(0)}=0$ or $\bar{y}_2^{(0)}=0$. If ${y}_2^{(0)}=0$ and $\bar{y}_2^{(0)}\neq0$ then the short-distance equations of motion read
\begin{equation}
\begin{split}
2a_{20}y_1^{(0)2}&=\left(\frac{r_V}{r}\right)^3,\\
a_{11}y_1^{(0)2}&=0,
\end{split}
\end{equation}
which imply that $a_{11}=0$. But then the condition $\bar{\Sigma}_2^{(0)}=0$ implies that all the cubic Galileon coefficients vanish and there can be no Vainshtein mechanism. If both ${y}_2^{(0)}=0$ and $\bar{y}_2^{(0)}=0$ then this condition implies that $a_{20}=0$ and $a_{02}=0$ (with $b_{02}$ unconstrained), and again there can be no Vainshtein mechanism. Finally, if ${y}_2^{(0)}\neq0$ and $\bar{y}_2^{(0)}=0$ then the condition $\bar{\Sigma}_2^{(0)}=0$ implies that $a_{20}=0$, $a_{11}=0$ and $a_{02}=0$ (while the short-distance equation of motion will require $b_{02}=0$), and once again there will be no Vainshtein screening of the Galileons.

\section{} \label{sec:app4}

In this section we show that, in bi-Galileon theory, the loophole mentioned at the end of Sec.\ \ref{sec:perturbations} leads to a contradiction after using the equations of motion. In the following we will use the same notation introduced in Appendix \ref{sec:app5}. At large distances the equations of motion are dominated by the linear functions $f_1^A(\bar{y}_1^{(0)},\bar{y}_2^{(0)})$, and the solutions are given by Eq.\ (\ref{eq:app4_linear_sols}). At short distances, on the other hand, the equations of motion are dominated by the cubic functions $f_3^A(y_1^{(0)},y_2^{(0)})$. Explicitly we have
\begin{equation}
\begin{split}
2f_3^1(y_1^{(0)},y_2^{(0)})&=\left(\frac{r_V}{r}\right)^3,\\
2f_3^2(y_1^{(0)},y_2^{(0)})&=0.
\end{split}
\end{equation}
The assumptions of the loophole case are that $\Sigma_3^{(0)}\equiv \Sigma_3(y_1^{(0)},y_2^{(0)})$ is singular positive semidefinite (but nonvanishing), and that $\bar{\Sigma}_3^{(0)}\equiv \Sigma_3(\bar{y}_1^{(0)},\bar{y}_2^{(0)})$ is singular negative semidefinite. For the matrix $\Sigma_3^{(0)}$ to be singular we have two options; one is that the rows of this matrix are nonzero but proportional to each other, and the other is that one of the rows is zero. The first option does not work because of the equations of motion, for if the second row is $\lambda$ times the first row (with $\lambda\neq0$), then
\begin{equation}
\begin{split}
f_3^2(y_1^{(0)},y_2^{(0)})&= \frac{1}{3}\left[y_1^{(0)}(\Sigma_3^{(0)})_{21}+y_2^{(0)}(\Sigma_3^{(0)})_{22}\right]\\
&=\frac{\lambda}{3}\left[y_1^{(0)}(\Sigma_3^{(0)})_{11}+y_2^{(0)}(\Sigma_3^{(0)})_{12}\right]\\
&=\lambda f_3^1(y_1^{(0)},y_2^{(0)}),
\end{split}
\end{equation}
where we have used the results of Appendix \ref{sec:app1}. We see that $f_3^2(y_1^{(0)},y_2^{(0)})=0$ (as implied by the second equation of motion) if and only if $f_3^1(y_1^{(0)},y_2^{(0)})=0$ also, contradicting the first equation of motion. The only possibility is to have the second row of the matrix $\Sigma_3^{(0)}$ equal to zero (the first row cannot be zero, again by the equation of motion and the assumption that $\Sigma_3^{(0)}$ is nonvanishing). With this choice the second equation of motion is satisfied as an identity.

To analyze this case, it is convenient to work with the ratios $\alpha\equiv y_1^{(0)}/y_2^{(0)}$ and $\beta\equiv \bar{y}_1^{(0)}/\bar{y}_2^{(0)}$ (this assumes $y_2^{(0)}\neq0$ and $\bar{y}_2^{(0)}\neq0$; we will come back to this assumption later). The matrices $\Sigma_3^{(0)}$ and $\bar{\Sigma}_3^{(0)}$ are then explicitly given by
\begin{equation}
\Sigma_3^{(0)}=\left( \begin{array}{cc} 3a_{30}\alpha^2+2a_{21}\alpha+a_{12} & a_{21}\alpha^2+2a_{12}\alpha+3a_{03} \\ 
a_{21}\alpha^2+2a_{12}\alpha+3a_{03} & a_{12}\alpha^2+6a_{03}\alpha+3b_{03} \end{array} \right)y_2^{(0)2},
\end{equation}
\begin{equation}
\bar{\Sigma}_3^{(0)}=\left( \begin{array}{cc} 3a_{30}\beta^2+2a_{21}\beta+a_{12} & a_{21}\beta^2+2a_{12}\beta+3a_{03} \\ 
a_{21}\beta^2+2a_{12}\beta+3a_{03} & a_{12}\beta^2+6a_{03}\beta+3b_{03} \end{array} \right)\bar{y}_2^{(0)2}.
\end{equation}
$\Sigma_3^{(0)}$ is singular positive semidefinite, with vanishing second row, if
\begin{equation} \label{eq:app4_cond1}
\begin{split}
3a_{30}\alpha^2+2a_{21}\alpha+a_{12}&>0,\\
a_{21}\alpha^2+2a_{12}\alpha+3a_{03}&=0,\\
a_{12}\alpha^2+6a_{03}\alpha+3b_{03}&=0.
\end{split}
\end{equation}
$\bar{\Sigma}_3^{(0)}$ is singular negative semidefinite if
\begin{equation} \label{eq:app4_cond2}
\begin{split}
(3a_{30}\beta^2+2a_{21}\beta+a_{12})(a_{12}\beta^2+6a_{03}\beta+3b_{03})-(a_{21}\beta^2+2a_{12}\beta+3a_{03})^2&=0,\\
(3a_{30}\beta^2+2a_{21}\beta+a_{12})+(a_{12}\beta^2+6a_{03}\beta+3b_{03})&<0.
\end{split}
\end{equation}
Notice that we must obviously have $\beta\neq \alpha$ to satisfy these conditions. Also, notice that $\beta\neq0$; this follows from the explicit expressions we have for the large-distance $\bar{y}_1^{(0)}$ and $\bar{y}_2^{(0)}$ (see Eqs.\ (\ref{eq:app4_linear_sols})), from where it follows that $\beta=-b_{01}/a_{01}$, which is nonzero since $b_{01}\neq0$ from the condition that the kinetic Lagrangian be strictly positive definite. We will next show that conditions (\ref{eq:app4_cond1}) and (\ref{eq:app4_cond2}) are inconsistent.

Suppose first that $(a_{12}\beta^2+6a_{03}\beta+3b_{03})$ is strictly negative (Eq.\ (\ref{eq:app4_cond2}) implies that it is nonpositive). Then we can solve for $3a_{30}$ from the first equation in (\ref{eq:app4_cond2}), finding that
\begin{equation} \label{eq:app4_1}
3a_{30}=\frac{1}{\beta^2}\left[\frac{(a_{21}\beta^2+2a_{12}\beta+3a_{03})^2}{(a_{12}\beta^2+6a_{03}\beta+3b_{03})}-(2a_{21}\beta+a_{12})\right].
\end{equation}
In addition, from Eqs.\ (\ref{eq:app4_cond1}) we have
\begin{equation}
\begin{split}
3a_{03}&=-a_{21}\alpha^2-2a_{12}\alpha,\\
3b_{03}&=-a_{12}\alpha^2-6a_{03}\alpha,
\end{split}
\end{equation}
from where we find that
\begin{equation} \label{eq:app4_2}
\begin{split}
(a_{12}\beta^2+6a_{03}\beta+3b_{03})&=(\beta-\alpha)(a_{12}(\beta+\alpha)+6a_{03}),\\
(a_{21}\beta^2+2a_{12}\beta+3a_{03})&=(\beta-\alpha)(a_{21}(\beta+\alpha)+2a_{12}).
\end{split}
\end{equation}
Using Eqs.\ (\ref{eq:app4_1}) and (\ref{eq:app4_2}), and performing some straightforward manipulations, we find
\begin{equation}
\begin{split}
(3a_{30}\alpha^2+2a_{21}\alpha+a_{12})&= \frac{\alpha^2}{\beta^2}\left[\frac{(\beta-\alpha)(a_{21}(\beta+\alpha)+2a_{12})^2}{(a_{12}(\beta+\alpha)+6a_{03})}-(2a_{21}\beta+a_{12})\right]+(2a_{21}\alpha+a_{12})\\
&= \frac{(\beta-\alpha)}{\beta^2}\left[\frac{\alpha^2(a_{21}(\beta+\alpha)+2a_{12})^2}{(a_{12}(\beta+\alpha)+6a_{03})}+(2a_{21}\alpha\beta+a_{12}(\beta+\alpha))\right]\\
&= \frac{(\beta-\alpha)^3}{\beta^2}\frac{(a_{21}\alpha+a_{12})^2}{(a_{12}(\beta+\alpha)+6a_{03})}\\
&= \frac{(\beta-\alpha)^4}{\beta^2}\frac{(a_{21}\alpha+a_{12})^2}{(a_{12}\beta^2+6a_{03}\beta+3b_{03})},\\
\end{split}
\end{equation}
where in the last line we used again Eq.\ (\ref{eq:app4_2}). Since we have assumed that $(a_{12}\beta^2+6a_{03}\beta+3b_{03})$ is strictly negative, we conclude that $(3a_{30}\alpha^2+2a_{21}\alpha+a_{12})$ must also be strictly negative, contradicting condition (\ref{eq:app4_cond1}).

Let us next assume that $(a_{12}\beta^2+6a_{03}\beta+3b_{03})=0$. Conditions (\ref{eq:app4_cond2}) then imply that also $(a_{21}\beta^2+2a_{12}\beta+3a_{03})=0$, and that $(3a_{30}\beta^2+2a_{21}\beta+a_{12})<0$. Taking these conditions together with (\ref{eq:app4_cond1}), we see that $\alpha$ and $\beta$ correspond to the two distinct roots of the polynomials
\begin{equation}
\begin{split}
p_1(x)&=a_{21}x^2+2a_{12}x+3a_{03}=a_{21}(x-\alpha)(x-\beta),\\
p_2(x)&=a_{12}x^2+6a_{03}x+3b_{03}=a_{12}(x-\alpha)(x-\beta).
\end{split}
\end{equation}
(This assumes implicitly that $a_{12}\neq0$ and $a_{21}\neq0$, for otherwise we cannot have $\alpha\neq\beta$ and satisfy the above conditions.) Since the two roots of the polynomials are the same, we can find the following relations between their coefficients:
\begin{equation}
a_{12}^2=3a_{21}a_{03},~~~~~~a_{12}a_{03}=a_{21}b_{03}.
\end{equation}
But if we now compute the discriminant of $p_1$, we find that it is given by $\Delta_1=4a_{12}^2-12a_{21}a_{03}=0$ from the above relations. One can similarly show that $\Delta_2$, the discriminant of $p_2$, also vanishes. This contradicts the fact that $\alpha\neq \beta$.

Finally, let us drop the assumptions we made at the beginning about $y_2^{(0)}$ and $\bar{y}_2^{(0)}$ being nonzero. First, if $y_2^{(0)}=0$, then the matrix $\Sigma_3^{(0)}$ simplifies to
\begin{equation}
\Sigma_3^{(0)}=\left( \begin{array}{cc} 3a_{30} & a_{21} \\ 
a_{21} & a_{12} \end{array} \right)y_1^{(0)2},
\end{equation}
and we have to set $a_{21}=0=a_{12}$ and require $a_{30}>0$ to have $\Sigma_3^{(0)}$ singular positive semidefinite and satisfy the equations of motion. The matrix $\bar{\Sigma}_3^{(0)}$ is then given by
\begin{equation}
\bar{\Sigma}_3^{(0)}=\left( \begin{array}{cc} 3a_{30}\bar{y}_1^{(0)2} & 3a_{03}\bar{y}_2^{(0)2} \\ 
3a_{03}\bar{y}_2^{(0)2} & 6a_{03}\bar{y}_1^{(0)}\bar{y}_2^{(0)}+3b_{03}\bar{y}_2^{(0)2} \end{array} \right).
\end{equation}
For $\bar{\Sigma}_3^{(0)}$ to be negative semidefinite, we need $3a_{30}\bar{y}_1^{(0)2}\leq0$, which contradicts the condition $a_{30}>0$ unless $\bar{y}_1^{(0)}=0$. But if this is the case then the equations of motion in the large-distance limit would imply that $b_{01}=0$; see Eq.\ (\ref{eq:app4_linear_sols}). This would contradict the requirement of having a strictly positive definite kinetic Lagrangian for the perturbations.

Next, let us assume $y_2^{(0)}\neq0$ but $\bar{y}_2^{(0)}=0$. The matrix $\bar{\Sigma}_3^{(0)}$ is then given by
\begin{equation}
\bar{\Sigma}_3^{(0)}=\left( \begin{array}{cc} 3a_{30} & a_{21} \\ 
a_{21} & a_{12} \end{array} \right)\bar{y}_1^{(0)2}.
\end{equation}
The condition that this matrix must be singular negative semidefinite implies
\begin{equation} \label{eq:app4_3}
\begin{split}
3a_{30}a_{12}&=a_{21}^2,\\
a_{30}&\leq0,\\
a_{12}&\leq0.
\end{split}
\end{equation}
Since $y_2^{(0)}\neq0$, as we assumed above, the Galileon coefficients must satisfy conditions (\ref{eq:app4_cond1}). If $a_{12}<0$, we can solve for $a_{30}$ from the first of Eqs.\ (\ref{eq:app4_3}), finding that
\begin{equation}
\begin{split}
3a_{30}\alpha^2+2a_{21}\alpha+a_{12}&= \frac{a_{21}^2}{a_{12}}\alpha^2+2a_{21}\alpha+a_{12}\\
&=\frac{1}{a_{12}}(a_{21}\alpha+a_{12})^2\leq0,
\end{split}
\end{equation}
contradicting conditions (\ref{eq:app4_cond1}). If $a_{12}=0$, then from (\ref{eq:app4_3}), $a_{21}=0$ also. The first equation in (\ref{eq:app4_cond1}) then reduces to $3a_{30}\alpha^2>0$. This in turn implies that $a_{30}>0$, contradicting (\ref{eq:app4_3}).


\end{document}